 \documentclass[preprint,review,12pt]{elsarticle}

\usepackage{epstopdf}
\usepackage{amssymb}
\usepackage{color}
\usepackage{amsmath}
\usepackage{mathtools}
\usepackage{xcolor,colortbl}
\usepackage{graphicx}
\definecolor{gray}{rgb}{0.82,0.82,0.82}

\usepackage{multirow}

\usepackage{multicol}
\usepackage[export]{adjustbox} 
\usepackage{geometry}

\journal{XXX}

\begin{document}

\begin{frontmatter}


\title{A five-point TENO scheme with adaptive dissipation based on a new scale sensor}

\author[a]{Haohan Huang\fnref{fnref1}}
\author[b]{Tian Liang\fnref{fnref1}}
\fntext[fnref1]{The first two authors contributed equally.}

\author[a,b,c,d]{Lin Fu\corref{cor1}}
\ead{linfu@ust.hk}
\cortext[cor1]{Corresponding author.}

\address[a]{Department of Mathematics, The Hong Kong University of Science and Technology, Clear Water Bay, Kowloon, Hong Kong}
\address[b]{Department of Mechanical and Aerospace Engineering, The Hong Kong University of Science and Technology, Clear Water Bay, Kowloon, Hong Kong}
\address[c]{HKUST Shenzhen-Hong Kong Collaborative Innovation Research Institute, Futian, Shenzhen, China}
\address[d]{Shenzhen Research Institute, The Hong Kong University of Science and Technology, Shenzhen, China}

\begin{abstract}
In this paper, a new five-point targeted essentially non-oscillatory (TENO) scheme with adaptive dissipation is proposed. With the standard TENO weighting strategy, the cut-off parameter $C_T$ determines the nonlinear numerical dissipation of the resultant TENO scheme. Moreover, according to the dissipation-adaptive TENO5-A scheme, the choice of the cut-off parameter $C_T$ highly depends on the effective scale sensor. However, the scale sensor in TENO5-A can only roughly detect the discontinuity locations instead of evaluating the local flow wavenumber as desired. In this work, a new five-point scale sensor, which can estimate the local flow wavenumber accurately, is proposed to further improve the performance of TENO5-A. In combination with a hyperbolic tangent function, the new scale sensor is  deployed to the TENO5-A framework for adapting the cut-off parameter $C_T$, i.e., the local nonlinear dissipation, according to the local flow wavenumber. Overall, sufficient numerical dissipation is generated to capture discontinuities, whereas a minimum amount of dissipation is delivered for better resolving the smooth flows. A set of benchmark cases is simulated to demonstrate the performance of the new TENO5-A scheme.
\end{abstract}


\end{frontmatter}
\section{Introduction} \label{sec1}
For hyperbolic conservation laws, one of the most difficult issues is the development of high-order numerical schemes with the capability of capturing discontinuities sharply and preserving the high-order accuracy in smooth regions. The essentially non-oscillatory (ENO) \cite{ENO} scheme has attracted lots of attention since it was proposed. Among a set of candidate fluxes, the ENO scheme selects the smoothest flux. Unlike ENO, the weighted ENO (WENO) scheme proposed by Liu et al. \cite{WENOL} uses a nonlinear convex combination of all candidate fluxes, including the non-smooth fluxes. This weighting strategy ensures the high-order accuracy in the smooth regions and the ENO property near discontinuities. After that, Jiang and Shu \cite{WENO} propose the WENO5-JS scheme by introducing a new smoothness indicator and a novel finite-difference framework. However, further investigation demonstrates that the accuracy order of the WENO5-JS scheme degenerates near the  critical points. To remedy this drawback, the WENO5-M \cite{WENOM} scheme remaps the weights calculated by WENO5-JS, and the WENO5-Z \cite{WENOZ} scheme introduces a new weighting strategy by employing the global smoothness indicator. In addition, WENO5-JS generally produces excessive numerical dissipation, which may smear the small-scale structures in the flow field. The WENO-Z+ \cite{WENOZ+} scheme enhances the contribution of the less smooth candidate stencil flux to reduce the numerical dissipation. Recently, Sun et al. \cite{MDCD1}\cite{MDCD2} devise a method to optimize a class of finite difference schemes with the Minimized Dispersion and Controllable Dissipation (MDCD) properties by two independent parameters. More recently, Sun et al. \cite{REN2} and Li et al. \cite{REN3} present a finite difference scheme with minimum dispersion and adaptive dissipation (MDAD) properties by establishing a correlation between the local wavenumber and numerical dissipation. Different from altering the coefficients of the background linear schemes, the fourth- and fifth-order weighted compact nonlinear schemes (WCNS) \cite{deng2000developing} are developed by employing the compact schemes as the background schemes. 
The Hermite WENO (HWENO) \cite{qiu2004hermite} schemes are proposed based on the Hermite polynomials. Benefiting from the compactness of the reconstruction in these schemes, the three-point reconstruction can generate a fifth-order accuracy scheme. Furthermore, Cai et al. \cite{cai2016positivity} apply the positivity-preserving techniques in the finite volume HWENO schemes for enhancing numerical stability. However, for the HWENO schemes, both the function values and the first-order derivatives need to be evolved in time and utilized in the reconstruction, which is nontrivial in terms of practical implementations. After that, Li et al. \cite{li2021multi} introduce the multi-resolution HWENO schemes that only reconstruct the function values and obtain the first-order derivatives by the high-order linear polynomials. Overall, the main drawback of the compact schemes is that a global tridiagonal matrix needs to be solved at each time step, rendering them less efficient. 
Other variants include the central WENO (CWENO) schemes \cite{levy1999central}\cite{levy2000compact}{\color{black}\cite{tsoutsanis2021arbitrary}\cite{tsoutsanis2023relaxed}\cite{tsoutsanis2021cweno}\cite{maltsev2023hybrid}}, the WENO-AO \cite{balsara2016efficient} and WENO-ZQ \cite{zhu2016new} schemes, the WENO scheme with automatic dissipation adjustment \cite{fernandez2021reduced}, and etc. 

Different from the WENO weighting strategy, Fu et al. \cite{TENO5}\cite{FU1}\cite{FU2}\cite{FU3}\cite{fu2021very} propose a family of TENO schemes for solving hyperbolic conservation laws. The TENO schemes introduce a threshold parameter $C_T$ to assess whether the contribution of one candidate stencil could be incorporated into the final flux computation. The benefit of this concept is that the TENO schemes can restore the background linear schemes exactly in the smooth regions. Nevertheless, the standard TENO schemes cannot deploy adaptive numerical dissipation in different regions, i.e., low numerical dissipation to resolve the high-wavenumber flows and sufficient numerical dissipation to capture discontinuities. Later, Fu et al. \cite{TENO5A}\cite{TENOA1} propose a series of TENO-A  schemes by adapting the threshold parameter $C_T$ based on a discontinuity sensor proposed by Ren et al. \cite{TENO5Asensor}.  However, the primary shortcoming of the discontinuity sensor is its inability to evaluate the local flow wavenumber accurately. Despite the fact that Su et al. \cite{REN1} and Sun et al. \cite{REN2} construct a six-point scale sensor that can assess the local flow wavenumber accurately, the sensor cannot be applied directly to the five-point scheme due to the limited number of available stencil points. In addition to adapting $C_T$ for the different flow scales, by replacing the polynomial reconstruction with a non-polynomial jump-like THINC reconstruction \cite{xiao2005simple}\cite{deng2022new}, the TENO5-THINC \cite{takagi2022novel} scheme deploys the standard TENO scheme in the smooth regions and the non-polynomial THINC reconstruction for resolving discontinuities based on a novel discontinuity-detection criterion. The performance of the TENO-family schemes has been extensively demonstrated for the compressible gas dynamics \cite{peng2021efficient}\cite{li2021low}\cite{fardipour2020development}\cite{tan2022two}\cite{meng2022targeted}\cite{hiejima2022high}\cite{ye2022alternative}, the multiphase flows  \cite{haimovich2017numerical}, the ideal magnetohydrodynamics (MHD) flows \cite{fu2019high}, the turbulent flows \cite{hamzehloo2021performance}\cite{lusher2020shock}\cite{motheau2020investigation}\cite{lusher2021assessment}\cite{di2021direct}\cite{gillespie2022shock} and the fluid-structure-acoustics interactions \cite{wang2020immersed}, etc. {\color{black} For more details about TENO schemes, please refer to \cite{fu2023review}.}

In this paper, a new five-point TENO scheme with adaptive dissipation is proposed by developing a novel five-point scale sensor. The main framework of the new scheme is based on the TENO5-A scheme. The numerical dissipation-related parameter $C_T$ is determined by the local flow wavenumber evaluated by the new five-point scale sensor. Additionally, a hyperbolic tangent function is employed to map the estimated wavenumber to a limited interval due to the unboundedness of the evaluated wavenumber. The new scheme achieves the adaptive dissipation control according to the local flow wavenumber, i.e., low dissipation is
deployed in the low wavenumber regions,  comparatively high dissipation is delivered in the high wavenumber regions, and adequate dissipation is generated for capturing  discontinuities. To demonstrate the performance of the new scheme, a set of 1D and 2D challenging benchmark cases with broadband flow length scales and shockwaves is simulated.

The remainder of the paper is organized as follows. 
(i) In section 2, the basic concept of the standard TENO5 scheme for scalar conservation law is briefly reviewed; (ii) In section 3, the new five-point scale sensor and the new TENO scheme are proposed in detail; (iii) In section 4, the performance of the new scheme is demonstrated by simulating a set of benchmark cases; (iv) Concluding remarks are given in the last section.

\section{Brief review of the TENO scheme}\label{sec2}

\begin{figure}[ht]
    \centering
    \setlength{\abovecaptionskip}{0.cm}
    \includegraphics[width=\linewidth]{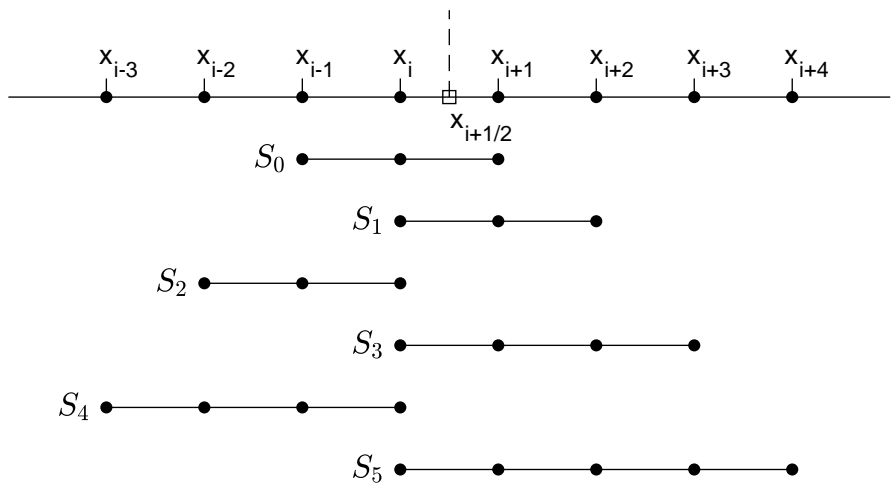}
    \caption{Candidate stencils of the high-order TENO scheme for reconstructing the cell interface flux at $i+\frac{1}{2}$ \cite{FU1}. The characteristic speed is assumed to be $\frac{\partial f(u)}{\partial u} > 0$. The stencil arrangement and the corresponding candidate stencil schemes for the scenario with $\frac{\partial f(u)}{\partial u} < 0$ can be obtained by symmetry at $i+\frac{1}{2}$.}
    \label{stencil}
\end{figure}

In the following sections, we consider the one-dimensional scalar hyperbolic conservation law
\begin{equation}
\label{eq_TT}
\frac{\partial u}{\partial t}+\frac{\partial f(u)}{\partial x} =0,
\end{equation}
where $u$ denotes the solution and $f$ is the flux function. For hyperbolic conservation laws, $\frac{\partial f(u)}{\partial u}$ denotes the characteristic signal speed. Without loss of generality, the characteristic speed is assumed to be $\frac{\partial f(u)}{\partial u} > 0$ in the following analysis. Then, a system of ordinary differential equations is formed by discretizing Eq.~(\ref{eq_TT}), 
\begin{equation}
\label{eq:ODE}
\frac{d u_{i}}{d t}=-\left.\frac{\partial f}{\partial x}\right|_{x=x_{i}}, \text{ } i=0, \cdots, N.
\end{equation}
$\left.\frac{\partial f}{\partial x}\right|_{x=x_{i}}$ can be approximated by a conservative finite-difference scheme as
\begin{equation}
    \left.\frac{\partial f}{\partial x}\right|_{x=x_{i}}=\frac{1}{\Delta x}(h_{i+1/2}-h_{i-1/2}),
\end{equation}
where $h(x)$ is an implicit function of $f(x)$, and defined as
\begin{equation}
f(x)=\frac{1}{\Delta x} \int_{x-\Delta x / 2}^{x+\Delta x / 2} h(\xi) d \xi .
\label{Eq:h_definition}
\end{equation}
Furthermore, $\left.\frac{\partial f}{\partial x}\right|_{x=x_{i}}$ can be numerically approximated following
\begin{equation}
\frac{d u_{i}}{d t} \approx-\frac{1}{\Delta x}\left(\widehat{f}_{i+1 / 2}-\widehat{f}_{i-1 / 2}\right),
\end{equation}
where $\widehat{f}_{i\pm 1/2}$ denotes the numerical flux and can be approximated by a convex combination of $K-2$ candidate-stencil fluxes,
\begin{equation}
\widehat{f}_{i+1 / 2}=\sum_{k=0}^{K-3} w_{k} \widehat{f}_{k, i+1 / 2}.
\end{equation}
The candidate stencil arrangement of the TENO scheme is shown in Fig.~\ref{stencil}. The TENO scheme employs a set of candidate stencils with incremental width and ensures that each candidate stencil contains at least one upwind point. The stencil width $r_k$ for the $K$-point scheme is summarized as 
\begin{equation}
\left\{r_{k}\right\}= \begin{cases}\{\underbrace{\left.3,3,3,4, \ldots, \frac{K+2}{2}\right\},}_{0, \ldots, K-3} & \text { if } \bmod (K, 2)=0, \\ \{\underbrace{3,3,3,4, \ldots, \frac{K+1}{2}}_{0, \ldots, K-3}\}, & \text { if } \bmod (K, 2)=1.\end{cases}
\end{equation}
A $(r_k-1)-$degree polynomial can be constructed corresponding to the candidate stencil $S_k$, as
\begin{equation}
h(x) \approx \hat{f}_{k}(x)=\sum_{l=0}^{r_{k}-1} a_{l, k} x^{l},
\end{equation}
where the coefficients $a_{l,k}$ are determined by satisfying Eq.~(\ref{Eq:h_definition}). The numerical flux at the cell interface $i+\frac{1}{2}$ can be approximated by the polynomial $\hat{f}_{k}(x)$ corresponding to each candidate stencil $S_k$. The smoothness measure of the $k$-th candidate stencil is defined as
\begin{equation}
    \gamma_k=(C+\frac{\tau_K}{\beta_{k,r_k}+\varepsilon})^q, \text{ } k=0,\dots,K-3,
\end{equation}
where $\varepsilon =10^{-40}$ to avoid the zero denominator. Different from the standard WENO5-JS scheme, for a stronger separation between different flow scales, $C=1$ and $q=6$ are adopted. Following \cite{WENO}, $\beta_{k,r_k}$ can be defined as 
\begin{equation}
\beta_{k, r_{k}}=\sum_{j=1}^{r_{k}-1} \Delta x^{2 j-1} \int_{x_{i-1 / 2}}^{x_{i+1 / 2}}\left(\frac{d^{j}}{d x^{j}} \hat{f}_{k}(x)\right)^{2} d x.
\label{smoothindicator}
\end{equation}
Then, to implement the ENO-like stencil selection strategy, the smoothness measure is normalized as
\begin{equation}
\chi_{k}=\frac{\gamma_{k}}{\sum_{i=0}^{K-3} \gamma_{i}} .
\end{equation}
Unlike the WENO schemes \cite{WENO}, TENO schemes either abandon the non-smooth stencils completely or apply the smooth ones with the optimal linear weights for the final  reconstruction. Specifically, a sharp cut-off function is defined as
\begin{equation}
\delta_{k}= \begin{cases}0, & \text { if } \chi_{k}<C_{T}, \\
1, & \text { otherwise, }\end{cases}
\end{equation}
where $C_T$ is a constant in the standard TENO schemes \cite{FU1}. It is noted that, instead of being a constant, $C_T$ can be adjusted dynamically to further control the numerical dissipation, as shown in the TENO-A schemes \cite{TENO5A}\cite{TENOA1}. At last, the final nonlinear weight of each candidate stencil can be computed by
\begin{equation}
    w_k=\frac{d_k\delta_k}{\sum_{i=0}^{K-3} d_i\delta_i}, \text{ } k=0,\dots,K-3,
\end{equation}
where $d_k$ represents the optimal weight of candidate stencil $S_k$. For obtaining the final $K-$th order scheme in the smooth regions, the values of $d_k$ are shown in Table~\ref{optimalweight}. Then, the final high-order reconstruction for the numerical flux at the cell interface $i+\frac{1}{2}$ is assembled as
\begin{equation}
\hat{f}_{i+1 / 2}^{K}=\sum_{k=0}^{K-3} w_{k} \hat{f}_{k, i+1 / 2}.
\end{equation}
\begin{table}[]
    \centering
    \setlength\tabcolsep{18pt}
    \renewcommand\arraystretch{0.8}
    \caption{Optimal weight $d_k$ of each candidate stencil for achieving the global $K-$th order scheme.}
    \begin{tabular}{llllllll}
    \hline
    Order & $d_{0}$ & $d_{1}$ & $d_{2}$ & $d_{3}$ & $d_{4}$ & $d_{5}$ &\\
    \hline\hline$K=3$ & 1 & & & & & \\
    $K=4$ & $\frac{3}{6}$ & $\frac{3}{6}$ & & & & \\
    $K=5$ & $\frac{6}{10}$ & $\frac{3}{10}$ & $\frac{1}{10}$ & & & \\
    $K=6$ & $\frac{9}{20}$ & $\frac{6}{20}$ & $\frac{1}{20}$ & $\frac{4}{20}$ & & & \\
    $K=7$ & $\frac{18}{35}$ & $\frac{9}{35}$ & $\frac{3}{35}$ & $\frac{4}{35}$ & $\frac{1}{35}$ &\\
    $K=8$ & $\frac{30}{70}$ & $\frac{18}{70}$ & $\frac{4}{70}$ & $\frac{12}{70}$ & $\frac{1}{70}$ & $\frac{5}{70}$ &\\
    \hline
    \end{tabular}
    \label{optimalweight}
\end{table}

Especially, for the standard TENO5 scheme, the candidate stencils involve $\{S_0, S_1, S_2\}$. With some algebraic derivations, the candidate numerical fluxes at the cell interface $i+\frac{1}{2}$ can be explicitly given by
\begin{equation}
\begin{aligned}
&\hat{f}_{0, i+1 / 2}=\frac{1}{6}\left(-f_{i-1}+5 f_{i}+2 f_{i+1}\right),\\
&\hat{f}_{1, i+1 / 2}=\frac{1}{6}\left(2 f_{i}+5 f_{i+1}-f_{i+2}\right),\\
&\hat{f}_{2, i+1 / 2}=\frac{1}{6}\left(2 f_{i-2}-7 f_{i-1}+11 f_{i}\right).
\end{aligned}
\end{equation}
Then, according to Eq.~(\ref{smoothindicator}), the explicit formulas for the smoothness indicators of the three candidate stencils are given as
\begin{equation}
\begin{aligned}
&\beta_{0}=\frac{1}{4}\left(f_{i-1}-f_{i+1}\right)^{2}+\frac{13}{12}\left(f_{i-1}-2 f_{i}+f_{i+1}\right)^{2},\\
&\beta_{1}=\frac{1}{4}\left(3 f_{i}-4 f_{i+1}+f_{i+2}\right)^{2}+\frac{13}{12}\left(f_{i}-2 f_{i+1}+f_{i+2}\right)^{2},\\
&\beta_{2}=\frac{1}{4}\left(f_{i-2}-4 f_{i-1}+3 f_{i}\right)^{2}+\frac{13}{12}\left(f_{i-2}-2 f_{i-1}+f_{i}\right)^{2}. 
\end{aligned}
\label{smoothnessindicator}
\end{equation}
Following \cite{WENOZ}\cite{FU1}, the global smoothness measure of the fifth-order TENO5 scheme is defined as $\tau_5=|\beta_1-\beta_2|$, and the cut-off parameter $C_T$ is set as $10^{-5}$ by spectral analysis.

{\color{black}
For the temporal integration of the resulting ODE Eq.~(\ref{eq:ODE}), the third-order strong-stability-preserving (SSP) Runge–Kutta scheme is utilized, which can be written as

\begin{equation}
\begin{aligned}
& u^{(1)}=u^n+\Delta t L\left(u^n\right), \\
& u^{(2)}=\frac{3}{4} u^n+\frac{1}{4} u^{(1)}+\frac{1}{4} \Delta t L\left(u^{(1)}\right), \\
& u^{n+1}=\frac{1}{3} u^n+\frac{2}{3} u^{(2)}+\frac{2}{3} \Delta t L\left(u^{(2)}\right).
\end{aligned}
\end{equation}
}

\section{The new five-point scale sensor and new TENO5-A scheme} \label{sec3}
This section is divided into two parts. We will first review the discontinuity sensor utilized in the standard TENO5-A scheme and propose a new five-point scale sensor. Then, a new TENO5-A scheme will be proposed by incorporating this new scale sensor.

\subsection{The new five-point scale sensor}
The discontinuity sensor $m$ deployed in the standard TENO5-A scheme is proposed by Ren et al. \cite{TENO5Asensor} as
\begin{equation}
    m=1-\min(1,\frac{\eta_{i+1/2}}{C_r}),
    \label{m}
\end{equation}
where
{\color{black}
\begin{equation}
    \eta_i=\frac{|2D f_{i+1/2}D f_{i-1/2}|+\varepsilon}{(D f_{i+1/2})^2+(D f_{i-1/2})^2+\varepsilon},\qquad \varepsilon=\frac{0.9C_r}{1-0.9C_r}\xi^2,\qquad Df_{i+1/2}=f_{i+1}-f_i,
    \label{g(m)}
\end{equation}}
and the corresponding parameters are $\xi=10^{-3}$, $C_r=0.24,$ and $\eta_{i+1/2}=\min(\eta_{i-1},\eta_i,\eta_{i+1})$. This sensor can roughly locate the discontinuities, but is incapable of  estimating the specific wavenumber. In the standard TENO5-A scheme, $C_T$ is adjusted dynamically to tailor the nonlinear numerical dissipation according to the smoothness of local flow scales as
\begin{equation}
\label{eq:CT_definition}
\left\{\begin{array}{l}
g(m) = {(1 - m)^4}(1 + 4m) ,\\
\overline{\beta}  =  {\alpha _1} - {\alpha _2}(1 - g(m)) ,\\
C_{T}=10^{-{\lfloor \bar{\beta}\rfloor}},
\end{array}\right.
\end{equation}
where $\left \lfloor  \right \rfloor$ is the Gauss bracket, $g(m)$ is a smoothing-kernel based mapping function, the parameter $\alpha _1 = 10.0$ and $\alpha _2 = 5.0$. When $m \approx 1$, $g(m) \approx 0$ and $C_T\approx10^{\alpha _2 - \alpha _1}$, which is typical for robust shock-capturing with strong nonlinear adaptation. When $m \approx 0$, $g(m) \approx 1$ and $C_T $ decreases to $ 10^{- \alpha _1}$, which is suitable for resolving the high-wavenumber physical fluctuations. With a proper choice of parameter $\alpha _1$ and $\alpha _2$, the TENO5-A scheme performs significantly better than the counterpart standard TENO5 scheme in terms of resolving the small-scale flow structures \cite{TENO5A}\cite{peng2021efficient}.

In this work, a novel five-point scale sensor will be proposed for evaluating the local flow wavenumber accurately to further enhance the performance of the TENO5-A scheme. According to \cite{REN3}, a six-point scale sensor is proposed based on the Taylor-series expansion, i.e.,
\begin{equation}
    f(x)-f(x_0)=\sum_{p=1}^{\infty}\frac{1}{p!}f^{(p)}(x_0)\Delta x^p,
\end{equation}
and for each order derivative, the variation of the solution is defined as  
\begin{equation}
    \Delta f_p=f^{(p)}(x_0)\Delta x^p.
\end{equation}
Since the lower order derivatives play more important roles in the
Taylor-series expansion than the higher order ones for smooth flow scales, it is reasonable to express the local scale of the solution by the ratio of $\Delta f_p$ at different orders as

\begin{equation}
K_{E S W}=\sqrt{\frac{\left|\Delta f_{3}\right|}{\left|\Delta f_{1}\right|}} \quad \text { or } \quad K_{E S W}=\sqrt{\frac{\left|\Delta f_{4}\right|}{\left|\Delta f_{2}\right|}}.
\label{eq_y1}
\end{equation}
The effectiveness of this scale sensor for estimating the scaled wavenumber can be analyzed as follows. Considering a pure sine function
\begin{equation}
    f(x)=A\sin (\omega x +\varphi),
\end{equation}
the derivatives of this function are given by
\begin{equation}
\begin{aligned}
&f^{(1)}(x)=A \omega \cos (\omega x+\varphi), \\
&f^{(2)}(x)=-A \omega^{2} \sin (\omega x+\varphi), \\
&f^{(3)}(x)=-A \omega^{3} \cos (\omega x+\varphi), \\
&f^{(4)}(x)=A \omega^{4} \sin (\omega x+\varphi),
\end{aligned}
\end{equation}
and the theoretical scaled wavenumber $K_{SW}$ is computed by
\begin{equation}
    K_{SW}=\omega \Delta x.
\end{equation}
Then, it can be straightforwardly deduced that
\begin{equation}
K_{S W}=\sqrt{\frac{\left|f^{(3)}\right|}{\left|f^{(1)}\right|}} \Delta x \quad \text { or } \quad K_{S W}=\sqrt{\frac{\left|f^{(4)}\right|}{\left|f^{(2)}\right|}} \Delta x,
\end{equation}
which is equal to
\begin{equation}
K_{S W}=\sqrt{\frac{\left|\Delta f_{3}\right|}{\left|\Delta f_{1}\right|}} \quad \text { or } \quad K_{S W}=\sqrt{\frac{\left|\Delta f_{4}\right|}{\left|\Delta f_{2}\right|}}.
\label{eq_y}
\end{equation}
Note that, when the two formulas are deployed separately, singular values may appear near the critical points and the inflection points. To deal with this problem, the two formulas in Eq.~(\ref{eq_y1}) are combined to achieve better performance for practical simulations as
\begin{equation}
K_{E S W}=\sqrt{\frac{\left|\Delta f_{3}\right|+\left|\Delta f_{4}\right|}{\left|\Delta f_{1}\right|+\left|\Delta f_{2}\right|+\varepsilon_1}},
\end{equation}
where $\varepsilon _1=10^{-12}$ denotes a small number for avoiding the zero denominator. {\color{black} Additionally, the final numerical results are not sensitive to the choice of $\varepsilon_1$ as long as it is a small value.}

Different from the six-point sensor developed in \cite{REN3}, due to insufficient stencil points in the five-point TENO5-A scheme, we propose to estimate each order derivative at $x_i$ instead of $x_{i+1/2}$. In practice, evaluating the wavenumber at $x_i$ or $x_{i+1/2}$ is almost identical. Finally, the five-point scale sensor can be written as
\begin{equation}
    K_{ESW}=\sqrt{\frac{\left|\Delta f_{3,i}\right|+\left|\Delta f_{4,i}\right|}{\left|\Delta f_{1,i}\right|+\left|\Delta f_{2,i}\right|+\varepsilon_1}},
\label{new}
\end{equation}
where
\begin{equation}
\begin{aligned}
\Delta f_{1, i} &=\frac{1}{12} f_{i-2}-\frac{2}{3} f_{i-1}+\frac{2}{3} f_{i+1}-\frac{1}{12} f_{i+2}, \\
\Delta f_{2, i} &=-\frac{1}{12} f_{i-2}+\frac{4}{3} f_{i-1}-\frac{5}{2} f_{i}+\frac{4}{3} f_{i+1}-\frac{1}{12} f_{i+2}, \\
\Delta f_{3, i} &=-\frac{1}{2} f_{i-2}+ f_{i-1}- f_{i+1}+\frac{1}{2} f_{i+2}, \\
\Delta f_{4, i} &= f_{i-2}-4 f_{i-1}+6f_{i}-4f_{i+1}+f_{i+2}.
\end{aligned}
\end{equation}
To demonstrate the performance of the newly proposed five-point scale sensor, a set of functions is considered to compare with the classical sensor in the standard TENO5-A scheme, i.e.,

 \noindent (a)
$$
f=\sin (20 \pi x), \quad-1 \leq x \leq 1;
$$
(b)
$$
f= \begin{cases}\sin (12 \pi x)-2, & -1 \leq x<0, \\ \sin \left(24.5 \pi x\right)+2, & 0 \leq x \leq 1;\end{cases}
$$
(c)
$$
f= \begin{cases}0, & -1 \leq x<0, \\ e^{x-1}\sin (32\pi x), & 0 \leq x \leq 1;\end{cases}
$$
(d)
$$
f=\sin (2\pi e^{x+1}x), \quad-1 \leq x \leq 1.
$$

\begin{figure}[htb]
    \centering
    \setlength{\abovecaptionskip}{0.cm}
    \includegraphics[width=\linewidth]{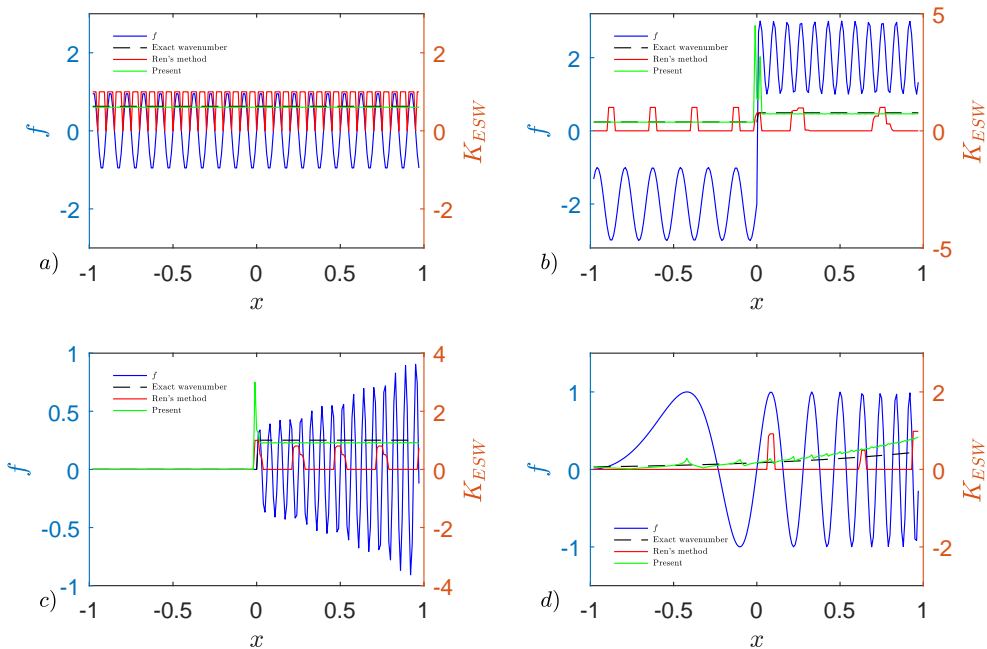}
    \caption{Functional value distributions of the Ren's sensor used in the standard TENO5-A scheme \cite{TENO5Asensor} and the newly proposed five-point scale sensor in this work.}
    \label{sensor}
\end{figure}

The computational results are shown in Fig.~\ref{sensor}. It can be seen that the wavenumber of critical points computed by Ren's method has distinct values from other points in the regions with the same theoretical wavenumber. Additionally, in case (b), the wavenumber calculated by Ren's method at the discontinuity is smaller than that at some critical points. In contrast, compared to the exact wavenumber distribution, the newly proposed five-point scale sensor can estimate the local flow wavenumber including that at critical points accurately. Moreover, the estimated wavenumber at discontinuities differs significantly from the other regions, which allows for deploying larger dissipation for sharp shock-capturing.

\subsection{The new five-point TENO-A scheme}

As mentioned above, the standard TENO5-A scheme determines the threshold parameter $C_T$ based on the discontinuity sensor $m$ in Eq.~(\ref{m}). Due to the unboundedness of the new scale sensor, an extra function, similar to $g(m)$ in TENO5-A, is needed to deploy the new scale sensor in the framework of TENO5-A. A hyperbolic tangent function is chosen due to its rigorous boundedness and good smoothness. The details of the new TENO5-A scheme are as follows.

\begin{figure}[htb]
    \centering
    \setlength{\abovecaptionskip}{0.cm}
    \includegraphics[width=\linewidth]{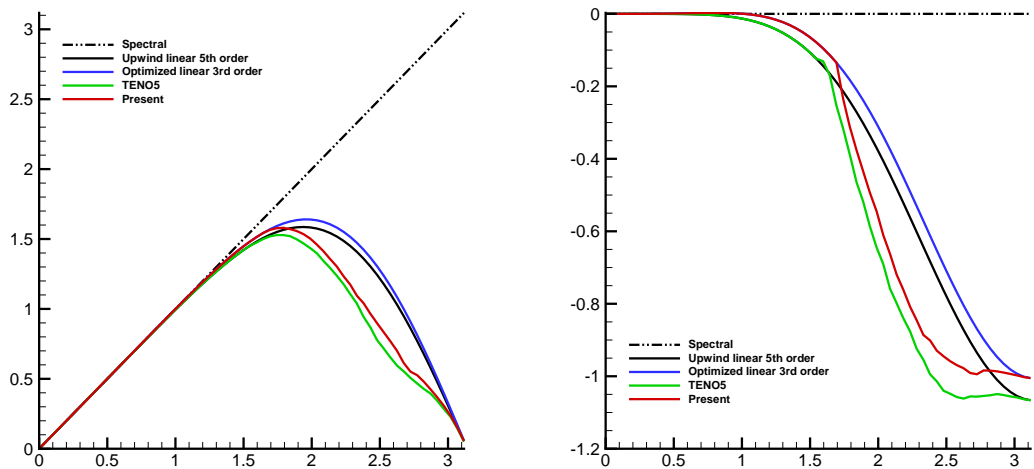}
    \caption{{\color{black}Dispersion (left) and dissipation (right) properties of the upwind 5th-order linear scheme, the optimized linear 3rd-order scheme, TENO5 scheme, and the present scheme, where the optimized linear 3rd-order scheme is from the combination of candidate stencils with the chosen optimized parameters $d_0=0.5065006634, d_1=0.3699651429,$ and $ d_2=0.1235341937$ \cite{TENO5A}.}}
    \label{spectral}
\end{figure}

Different from the standard TENO5 scheme \cite{TENO5}, the sixth-order global smoothness measure $\tau_5$ in the weighting strategy is taken as \cite{TENO5A}\cite{FU3}
\begin{equation}
\begin{aligned}
\tau_5 &=\frac{1}{5040}\left|5788 f_{i-2}^{2}+f_{i-2}\left(-45681 f_{i-1}+64843 f_{i}-38947 f_{i+1}+8209 f_{i+2}\right)+f_{i-1}\left(93483 f_{i-1}\right.\right.\\
&\left. -275836f_{i}+173498 f_{i+1}-38947 f_{i+2}\right)+f_{i}\left(210993 f_{i}-275836 f_{i+1}+64843 f_{i+2}\right) \\
&\left.+f_{i+1}\left(93483 f_{i+1}-45681 f_{i+2}\right)+5788 f_{i+2}^2\right|.
\end{aligned}
\end{equation}
For adaptive numerical dissipation, $C_T$ is determined by the newly proposed five-point scale sensor, 
\begin{equation}
\left\{\begin{array}{l}
g(K_{ESW})=\tanh(1.01K_{ESW}), \\
\bar{\beta}=\alpha_{1}-\alpha_{2}(g(K_{ESW})), \\
C_{T}=10^{-{\lfloor\bar{\beta}\rfloor}},
\end{array}\right.
\end{equation}
where $K_{ESW}$ is computed by Eq.~(\ref{new}), the parameter $\alpha_1=10$ and $\alpha_2=5$. Here, the function $\tanh(\cdot)$ is introduced for achieving the boundedness between $0$ and $1$. Then, the cut-off function is similarly defined as
\begin{equation}
\delta_{k}= \begin{cases}0, & \text { if } \chi_{k}<C_{T}, \\
1, & \text { otherwise. }\end{cases}
\end{equation}
At last, the final weight of each candidate stencil is given by 
\begin{equation}
    w_k=\frac{d_k\delta_k}{\sum_{i=0}^2 d_i\delta_i}, \text{ } k=0,1,2,
\end{equation}
where the optimal linear weights are optimized by spectral analysis, and given as $d_0=0.5065006634, d_1=0.3699651429,$ and $ d_2=0.1235341937$ \cite{TENO5A}.
The left formulas are the same as the standard TENO5-A scheme and are not shown here for brevity. {\color{black} Additionally, the dispersion and dissipation property of the present scheme can be analysed by the approximated dispersion relation (ADR) analysis \cite{pirozzoli2006spectral}\cite{zhao2019general}. As shown in Fig.~\ref{spectral}, both the dispersion and dissipation properties of the present scheme are better than TENO5. And the optimal background linear scheme can be restored exactly up to an intermediate wavenumber.}
\section{Numerical validation}\label{sec4}
In this section, a set of benchmark cases is simulated by WENO5-Z, TENO5, TENO5-A, and the present five-point scheme to access the performance.  The ideal gas equation $p=(\gamma-1)\rho e$ with $\gamma=1.4$ is employed to close the Euler equations. Unless otherwise specified, the CFL number is set as 0.4, the Rusanov scheme \cite{Rusanov1961} is used for flux splitting, and the Roe average is utilized for characteristic decomposition. In terms of the time integration, the third-order strong-stability-preserving (SSP) Runge–Kutta \cite{gottlieb2001strong} scheme is chosen as default. All the grids used in this section are Cartesian and uniform.  

\subsection{Accuracy test}

Considering the linear advection problem with a smooth initial condition to verify the accuracy order of the proposed scheme in the smooth regions, the governing equation and the initial condition are given as
\begin{equation}
\frac{\partial u}{\partial t}+\frac{\partial u}{\partial x}=0, \text{ } u(x,0)=\sin (\pi x), \text{ } 0\leq x \leq 2.
\end{equation}
The grid resolution is chosen as $N=20,40,80,160,320$, and $640$, respectively. $\Delta t$ is set as $\Delta x^{\frac{p}{3}}$, where $p$ is the theoretical accuracy order. 

The statistics of the $L_1$ and $L_\infty$ errors and the corresponding accuracy order are shown in Table~\ref{table1}. As expected, TENO5-A and the present scheme show third-order convergence. This case verifies that both TENO5-A and the newly proposed scheme can restore the optimal background linear scheme in the smooth regions, and the accuracy order of the optimal background linear scheme is third with the given optimal linear weights in the TENO5-A scheme \cite{TENO5A}. {\color{black} Since the TENO schemes either abandon the non-smooth stencils completely or apply the smooth ones with the optimal linear weights for the final reconstruction, meanwhile, this case only contains smooth functions, TENO5-A and the newly proposed scheme both restore the optimal background linear scheme exactly. Therefore, TENO5-A and the newly proposed scheme have the same performance in this case.}

\begin{table}[htb]
\centering
\setlength\tabcolsep{15pt}
\renewcommand\arraystretch{0.8}
\caption{The error statistics and the corresponding accuracy orders of TENO5-A and the present scheme.}
\begin{tabular}{llllll}
\hline \hline
Scheme                       & $N$   & $L_1$ error & $L_1$ order & $L_\infty$ error & $L_\infty$ order \\ \hline \hline
\multirow{5}{*}{TENO5-A}     & 20  & 2.03E-03   & -          & 3.28E-03     &      -        \\
                             & 40  & 2.86E-04   & 2.83       & 4.58E-04     & 2.84         \\
                             & 80  & 3.71E-05   & 2.95       & 5.89E-05     & 2.96         \\
                             & 160  & 4.69E-06   & 2.98       & 7.41E-06     & 2.99         \\ 
                             & 320 & 5.89E-07 & 2.99 & 9.27E-07 & 3.00\\
                             & 640 & 7.37E-08 & 3.00 & 1.16E-07 & 3.00\\
                             \hline
\multirow{5}{*}{Present}       & 20  & 2.03E-03  & -          & 3.28E-03     & -            \\
                             & 40  & 2.86E-04   & 2.83       & 4.58E-04     & 2.84         \\
                             & 80  & 3.71E-05   & 2.95       & 5.89E-05     & 2.96         \\
                             & 160  & 4.69E-06   & 2.98       & 7.41E-06     & 2.99   \\
                             & 320 & 5.89E-07 & 2.99 & 9.27E-07 & 3.00\\
                             & 640 & 7.37E-08 & 3.00 & 1.16E-07 & 3.00
                             
                             \\ \hline\hline

\end{tabular}
\label{table1}
\end{table}
\subsection{Linear advection of multiple waves}
This case is taken from \cite{WENO} and we solve the linear advection equation
\begin{equation}
    \frac{\partial u}{\partial t}+\frac{\partial u}{\partial x}=0,
\end{equation}
with the initial condition given as
\begin{equation}
u(x, 0)=\left\{\begin{array}{cc}
\frac{1}{6}[G(x-1, \beta, z-\theta)+G(x-1, \beta, z+\theta)+4 G(x-1, \beta, z)], & \text { if } 0.2 \leq x<0.4, \\
1, & \text { if } 0.6 \leq x \leq 0.8, \\
1-|10(x-1.1)|, & \text { if } 1.0 \leq x \leq 1.2, \\
\frac{1}{6}[F(x-1, \alpha, a-\theta)+F(x-1, \alpha, a+\theta)+4 F(x-1, \alpha, a)], & \text { if } 1.4 \leq x<1.6, \\
0, & \text { otherwise, }
\label{icc}
\end{array}\right.    
\end{equation}
where
\begin{equation}
G(x, \beta, z)=e^{-\beta(x-z)^{2}}, F(x, \alpha, a)=\sqrt{\max \left(1-\alpha^{2}(x-a)^{2}, 0\right)} .  \label{ic1}  
\end{equation}
The parameters in Eq.~(\ref{icc}) and Eq.~(\ref{ic1}) are
\begin{equation}
a=0.5, z=-0.7, \theta=0.005, \alpha=10, \beta=\frac{\log 2}{36 \theta^{2}} .    
\end{equation}
The initial condition consists of a Gaussian pulse, a square wave, a sharp triangle wave, and a half ellipse arranged from the left to the right in the computational domain $x\in [0,2]$. The equation is solved by a uniform grid with $N=200$, and the final evolution time is $t=2$ and $18$, respectively. The exact solution is the theoretical solution of the linear advection equation with a constant speed of propagation.

\begin{figure}[htb]
    \centering
    \setlength{\abovecaptionskip}{0.cm}
    \includegraphics[width=0.8\linewidth]{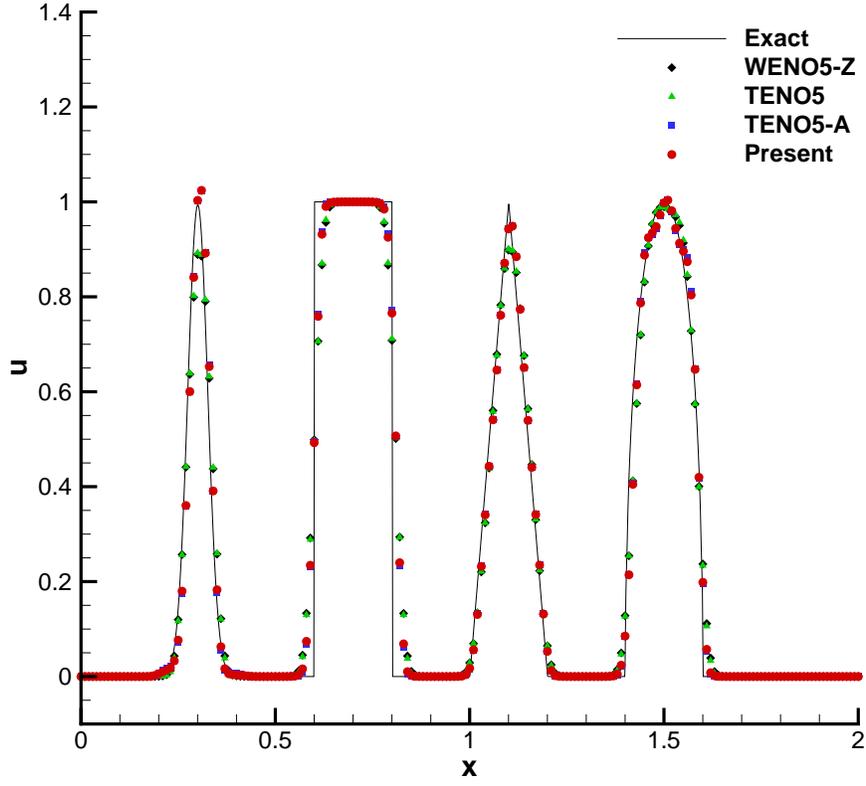}
    \caption{{\color{black}Linear advection of multiple waves: $u$ distributions from WENO5-Z, TENO5, TENO5-A, and the present scheme at the simulation time $t=2$. The spatial discretization is on 200 uniform grid points. ``Exact'' denotes the theoretical solution of the linear advection equation with the constant speed of propagation.}}
    \label{linearadvection}
\end{figure}

\begin{figure}[htb]
    \centering
    \setlength{\abovecaptionskip}{0.cm}
    \includegraphics[width=0.8\linewidth]{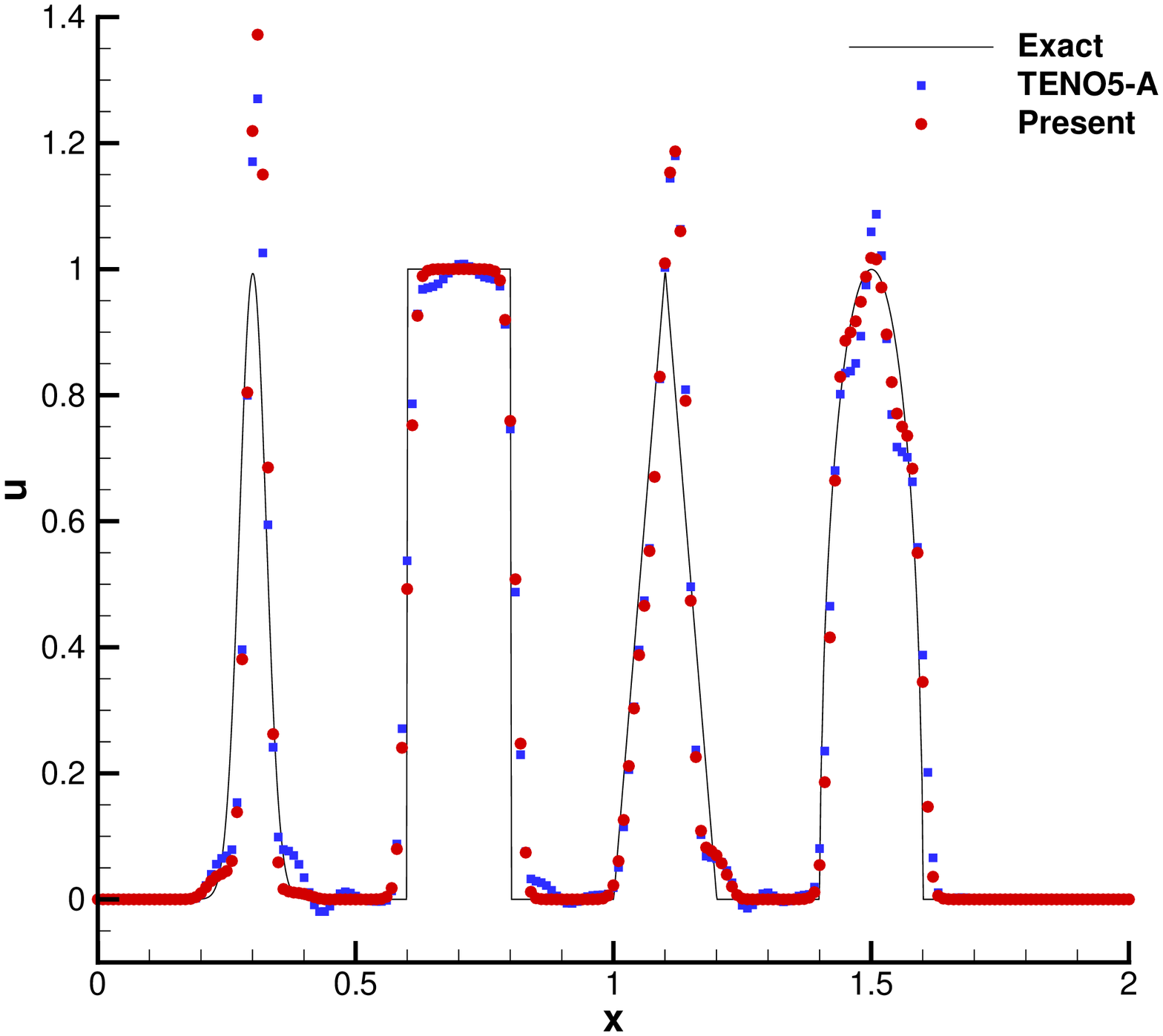}
    \caption{{\color{black}Linear advection of multiple waves: $u$ distributions from TENO5-A and the present scheme at the simulation time $t=18$. The spatial discretization is on 200 uniform grid points. ``Exact'' denotes the theoretical solution of the linear advection equation with the constant speed of propagation. The numerical error accumulates substantially after a long time evolution.}}
    \label{linearlong}
\end{figure}

As shown in Fig.~\ref{linearadvection}, with the short time evolution,  the performance of TENO5-A and the present scheme does not have obvious differences. However, for the Gaussian pulse and the sharp triangle wave, both TENO5-A and the present scheme capture them more sharply than WENO5-Z and TENO5. In terms of the long-time evolution where the numerical error accumulates substantially, as shown in Fig.~\ref{linearlong}, the result of TENO5-A has obvious oscillations even in the smooth regions. The present scheme exhibits notable advantages in preserving the overall shape and capturing the square wave sharply.
\subsection{Shock-tube problem}\label{1deulerbegin}
{\color{black}
From section \ref{1deulerbegin} to section \ref{1deulerend}, the one-dimensional Euler equations are solved, which can be written as
\begin{equation}
\left(\begin{array}{c}
\rho \\
\rho u \\
E
\end{array}\right)_t +\left(\begin{array}{c}
\rho u \\
\rho u^2+p \\
u(E+p)
\end{array}\right)_x=0,
\end{equation}
}
{\color{black} where $\rho$ denotes the density, $u$ is the velocity, $E$ is the total energy, and $p$ is the pressure.}

In this section, two typical shock-tube problems are solved to validate the shock-capturing capability of the proposed scheme. 

The initial condition for the Sod's problem \cite{Sod1978} is
\begin{equation}
(\rho, u, p)= \begin{cases}(1,0,1), & \text{ if } 0 \leq x <0.5, \\ (0.125,0,0.1), & \text{ if } 0.5 \leq x \leq 1,\end{cases}
\end{equation}
and the final simulation time is set as $t=0.2$.

The initial condition for the Lax's problem \cite{Lax1954} is
\begin{equation}
(\rho, u, p)= \begin{cases}(0.445,0.698,3.528), & \text{ if } 0 \leq x <0.5, \\ (0.5,0,0.5710), & \text{ if } 0.5 \leq x \leq 1,\end{cases}
\end{equation}
and the final simulation time is set as $t=0.14$.

\begin{figure}[htbp]
    \centering
    \setlength{\abovecaptionskip}{0.cm}
    \includegraphics[width=\linewidth]{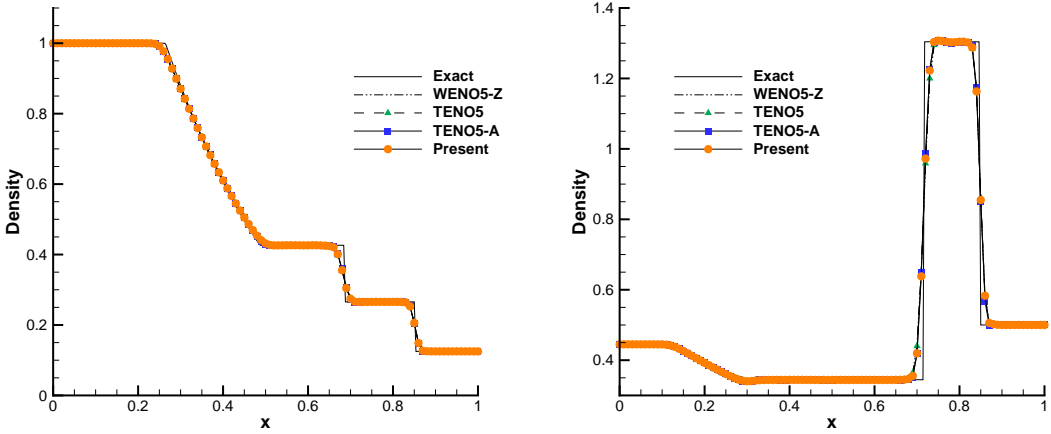}
    \caption{{\color{black}Shock-tube problem: the computed density distributions of the Sod's problem (left) and the computed density distributions of the Lax's problem (right). Discretization is both on 100 uniform grid points. ``Exact” both denotes the theoretical solution of the corresponding Riemann problem.}}
    \label{sod}
\end{figure}

These two cases are solved by WENO5-Z, TENO5, TENO5-A, and the present scheme with the resolution of $N=100$. The results are shown in Fig.~\ref{sod}. The ``Exact” reference is the theoretical solution of the corresponding Riemann problem. The results show that all these schemes considered can capture the discontinuities without artificial oscillations. 

\subsection{Interacting blast waves}
The two-blast-wave interaction taken from Woodward and Colella \cite{Woodward1984} is considered. The initial condition is
\begin{equation}
(\rho, u, p)= \begin{cases}(1,0,1000), & \text { if } 0 \leq x<0.1, \\ (1,0,0.01), & \text { if } 0.1 \leq x<0.9, \\ (1,0,100), & \text { if } 0.9 \leq x \leq 1.\end{cases}
\end{equation}
The reflective boundary condition is used at $x=0$ and $x=1$. Meanwhile, it is solved by a uniform grid with $N=400$ and the final evolution time $t=0.038$. The Roe scheme with entropy-fix is utilized for the numerical flux splitting, and the CFL number is set as 0.35 for good robustness. The exact solution is solved by the WENO5-JS scheme with $N=2000$. 

\begin{figure}[htbp]
    \centering
    \setlength{\abovecaptionskip}{0.cm}
    \includegraphics[width=\linewidth]{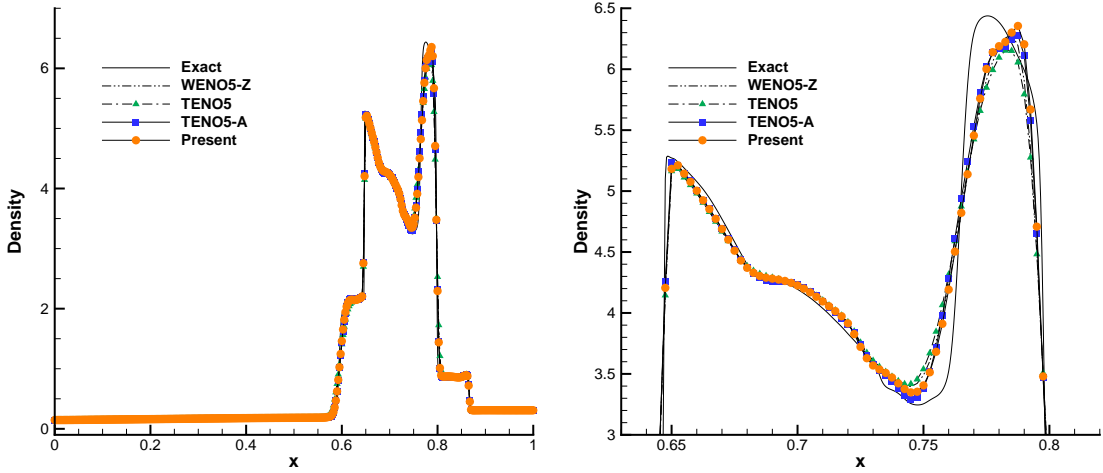}
    \caption{{\color{black}Interacting blast waves: the computed density distributions from various schemes (left) and the zoomed-in view (right). Discretization is on 400 uniform grid points. ``Exact” denotes the solution solved by WENO5-JS with 2000 grid points.}}
    \label{blast}
\end{figure}

As shown in Fig.~\ref{blast},  near the density peak at $x=0.78$, TENO5-A and the present scheme perform better than WENO5-Z and TENO5.

\subsection{Shock–density wave interaction}\label{1deulerend}

This case is taken from Shu and Osher \cite{shuosher}. The initial condition is 
\begin{equation}
(\rho, u, p)= \begin{cases}(3.857,2.629,10.333), & \text { if } 0 \leq x<1, \\ (1+0.2\sin (5(x-5)),0,1), & \text { if } 1 \leq x<10.\end{cases}
\end{equation}
This case is designed by simulating a Mach 3 shock interacting with a sine wave. The computed density $\rho$ is plotted at $t=1.8$ with $N=200$, and the exact solution is solved by the WENO5-JS scheme with $N=2000$. 

\begin{figure}[htbp]
    \centering
    \setlength{\abovecaptionskip}{0.cm}
    \includegraphics[width=\linewidth]{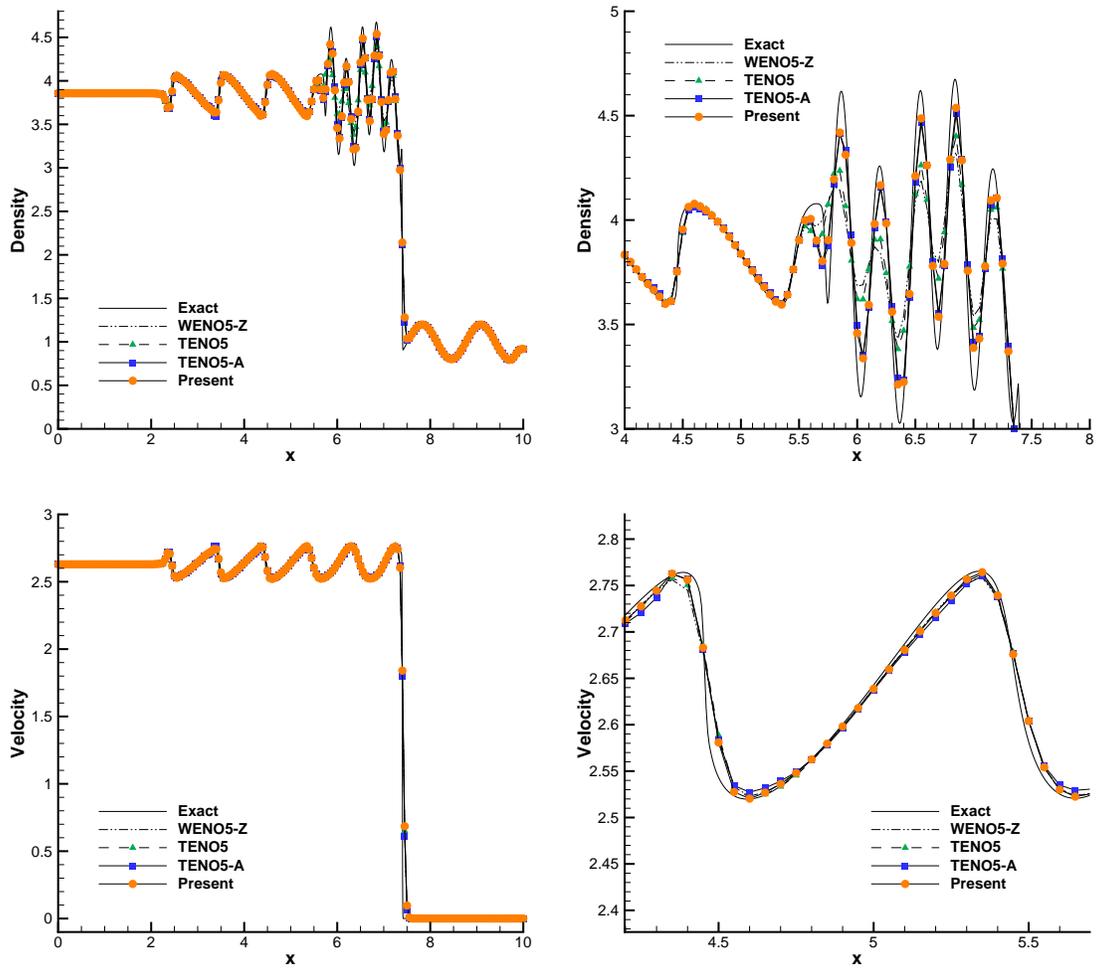}
    \caption{{\color{black}Shock–density wave interaction: solutions from various schemes. Top: the computed density distributions (left) and the zoomed-in view (right). Bottom: the computed velocity distributions (left) and the zoomed-in view (right). Discretization is on 200 uniform grid points. ``Exact” denotes the solution solved by WENO5-JS with 2000 grid points.}}
    \label{shuosher}
\end{figure}

As shown in Fig.~\ref{shuosher}, all the considered schemes are capable of capturing the acoustic waves. In terms of the density distribution, TENO5-A and the present scheme show a higher resolution than WENO5-Z and TENO5 in resolving the high-wavenumber physical fluctuations. The standard TENO5-A scheme also generates slight overshoots around $x=3.25$. For the computed velocity distribution, the present scheme features the best resolution in maintaining the amplitude of the velocity profile from $x=4.3$ to $5.6$. Both the zoomed-in views show that the present scheme has lower numerical dissipation than the other schemes.

\subsection{Double Mach reflection of a strong shock}
{\color{black}
In this section, the two-dimensional Euler equations are solved, which can be written as
\begin{equation}
\left(\begin{array}{c}
\rho \\
\rho u \\
\rho v \\
E
\end{array}\right)_t+\left(\begin{array}{c}
\rho u \\
\rho u^2+p \\
\rho u v \\
u(E+p)
\end{array}\right)_x+\left(\begin{array}{c}
\rho v \\
\rho u v \\
\rho v^2+p \\
v(E+p)
\end{array}\right)_y =0,
\end{equation}
where $u$ and $v$ denote the velocity along $x$- and $y$-direction, respectively.
}

The initial condition is given as
\begin{equation}
(\rho, u, v, p)= \begin{cases}(1.4,0,0,1), & \text { if } y < 1.732(x-0.1667), \\ (8,7.145,-4.125,116.8333), & \text{ otherwise.}\end{cases}
\end{equation}
The initial condition describes a Mach 10 shockwave moving from left to right with an incidence angle of $60^{\circ}$. As for the boundary condition, the inflow and outflow boundary conditions are implemented for the left and right sides of the computational domain, respectively. For the top side, the boundary condition follows the exact solution of a Mach 10 moving shockwave. In terms of the bottom side, the boundary condition in the region from $x=0$ to $x=0.1667$ follows the post-shock condition, whereas that in the remaining region from $x=0.1667$ to $x=4.0$ follows the reflective wall condition. The final evolution time is $t=0.2$, and the grid resolution is $1024\times 256$. 

\begin{figure}[htb]
    \centering
    \setlength{\abovecaptionskip}{0.cm}
    \includegraphics[width=0.85\linewidth]{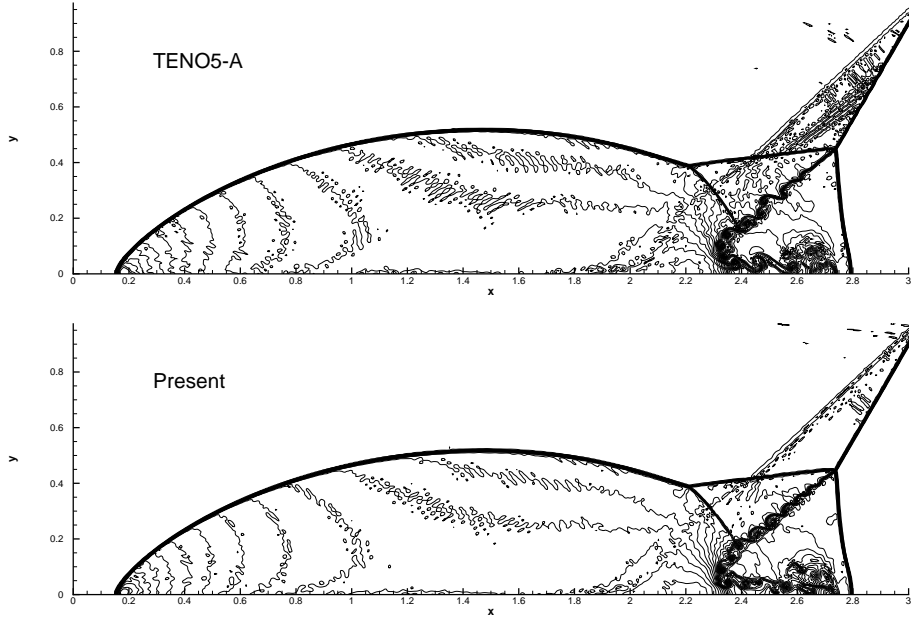}
    \caption{Double Mach reflection of a strong shock: density contours computed from TENO5-A and the present scheme. Both plots are drawn with 43 contourlines between 1.887 and 20.9. The resolution is $1024\times 256$.}
    \label{doubleb}
\end{figure}

\begin{figure}[htbp]
    \centering
    \setlength{\abovecaptionskip}{0.cm}
    \includegraphics[width=\linewidth]{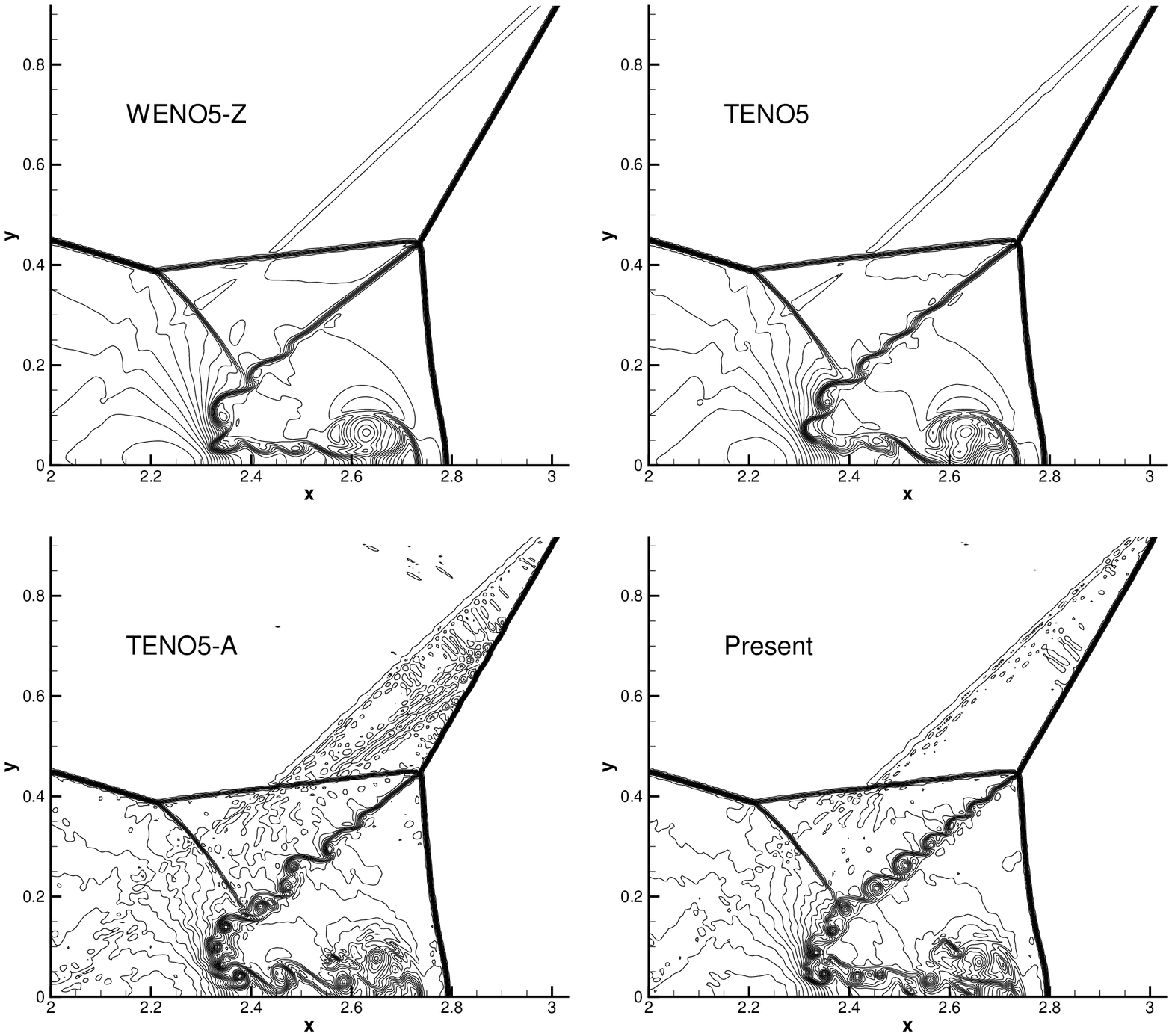}
    \caption{Same as Fig.~\ref{doubleb}, but the zoomed-in view from various schemes.}
    \label{double}
\end{figure}

This result of this case is very sensitive to the dissipation and dispersion properties of the deployed numerical scheme. Fig.~\ref{doubleb} plots the density contours from TENO5-A and the present scheme. Overall speaking, the present scheme generates much less numerical noise than TENO5-A behind the incident moving shockwave. Fig.~\ref{double} only presents the region of $[2,3]\times[0,0.9]$ for the ease of comparison. Each scheme exhibits a significantly different performance. The present scheme resolves the richest small-scale structures, indicating the lowest built-in numerical dissipation. Besides, TENO5-A generates too much numerical noise and fewer small-scale structures than the new scheme. Unlike the above two schemes,  both WENO5-Z and TENO5 produce excessive numerical dissipation, but TENO5 still performs better than WENO5-Z.

{\color{black}

In order to demonstrate the computational efficiency of the present scheme, table~\ref{cost_dm} shows the statistics of computational costs with WENO5-Z, TENO5, TENO5-A, and the present scheme. It can be seen that the computational effort is approximately the same for TENO5-A and the present scheme.}
\begin{table}[hbt]
	\centering
	\begin{tabular}{lccccc}
			\hline
			& WENO5-Z & TENO5 & TENO5-A  & Present  \\
			\hline
			 $512 \times 128$ [s] & 54.99  & 55.10  & 78.47 & 80.41 \\
			\hline
		\end{tabular}
	\caption{{\color{black}The computational time statistics with different numerical schemes.}}
	\label{cost_dm}
\end{table}

\subsection{Rayleigh-Taylor instability}
The inviscid Rayleigh-Taylor instability case proposed by Xu and Shu \cite{Xu2005} is considered, {\color{black} where the two-dimensional Euler equations with gravity are solved, i.e.,
\begin{equation}
\left(\begin{array}{c}
\rho \\
\rho u \\
\rho v \\
E
\end{array}\right)_t+\left(\begin{array}{c}
\rho u \\
\rho u^2+p \\
\rho u v \\
u(E+p)
\end{array}\right)_x+\left(\begin{array}{c}
\rho v \\
\rho u v \\
\rho v^2+p \\
v(E+p)
\end{array}\right)_y =\left(\begin{array}{c}
0 \\
0 \\
\rho g \\
\rho v g
\end{array}\right),
\end{equation}
here, $g = 1$ denotes the gravity.
}

and the initial condition is
\begin{equation}
(\rho, u, v, p)= \begin{cases}(2,0,-0.025c\cos (8\pi x),1+2y), & \text { if } 0\leq y < 0.5, \\ (1,0,-0.025c\cos (8\pi x),y+1.5), & \text{ if } 0.5\leq y \leq 1,\end{cases}
\end{equation}
where $c=\sqrt{\frac{\gamma p}{\rho}}$ is the sound speed with $\gamma=\frac{5}{3}$, and the computational domain is defined as $[0,0.25]\times[0,1]$. For the left and right sides of the computational domain, the reflective boundary condition is enforced. In terms of the bottom and top sides, constant primitive variables $(\rho , u,v,p)=(2,0,0,1)$ and $(\rho , u,v,p)=(1,0,0,2.5)$ are imposed, respectively. The final evolution time is $t=1.95$ and the gird resolution is $96\times 384$. Especially, the Roe scheme is used for flux splitting.

\begin{figure}[htbp]
    \centering
    \setlength{\abovecaptionskip}{0.cm}
    \includegraphics[width=0.8\linewidth]{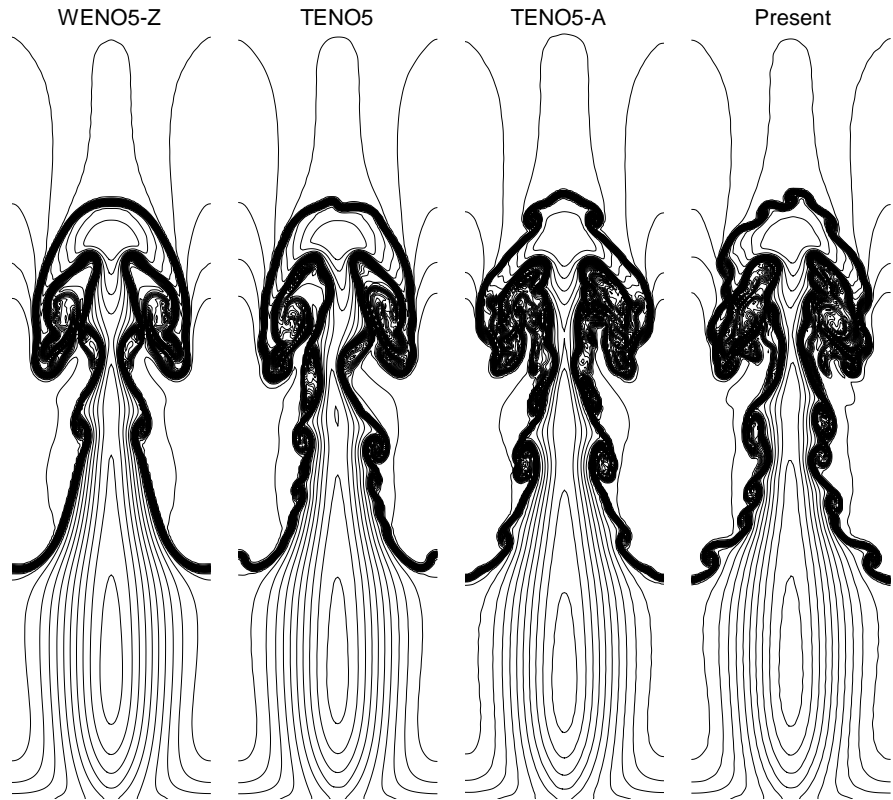}
    \caption{Rayleigh-Taylor instability: density contours computed from various schemes. All plots are drawn with 43 contourlines between 0.9 and 2.2. The resolution is $96 \times 384$.}
    \label{ra}
\end{figure}

As shown in Fig.~\ref{ra}, the present scheme resolves much richer vortical structures than the other schemes, indicating its low numerical dissipation. It is noted that the solutions from TENO5, TENO5-A, and the present scheme are not symmetric, which is mainly due to the fact that the low numerical dissipation cannot suppress the numerical disturbances from the machine round-off errors \cite{fleischmann2019numerical}\cite{wakimura2022symmetry}. 

\section{Conclusions}\label{sec5}
In this paper, a five-point TENO scheme with adaptive dissipation based on a new scale sensor is proposed for solving hyperbolic conservation laws. Compared to the discontinuity sensor in the standard TENO5-A scheme, the new scale sensor is capable of evaluating the local flow wavenumber. Benefiting from that, the proposed scheme adapts its dissipation according to the local flow wavenumber. A set of benchmark cases is simulated, and the performance of the proposed scheme is summarized as follows.

The proposed scheme can restore the optimal background linear scheme in the smooth regions without degeneration. The proposed scheme exhibits overall less numerical dissipation than the standard TENO5-A scheme while preserving the discontinuity more sharply. In the case of ``double Mach reflection of a strong shock”, the proposed scheme resolves more small-scale structures in the blow-up region and generates significantly less numerical noise than TENO5-A; In the case of the long-time evolution of the multiple waves, the advantage of the proposed scheme is obvious in terms of suppressing the artificial numerical oscillations.

The proposed scheme is easy to be implemented into an existing code since the same stencil points are utilized as the classical WENO5-JS scheme. And it has much lower dissipation than the existing five-point shock-capturing schemes. The present idea can also be extended for the very high-order TENO schemes with adaptive dissipation control. {\color{black} Our future work will also deploy the proposed schemes in more complicated simulations, including the MHD and multiphase flows. It is further noted that we do not see obvious barriers to extend the present scheme to turbulent flow simulations, and the relevant work will be reported in a separate forthcoming paper.}

\section*{Declaration of Competing Interest}

The authors declare that they have no known competing financial interests or personal relationships that could have appeared to influence the work reported in this paper. 

\section*{Data availability}
The data that support the findings of this study are available on request from the corresponding author, LF.

\section*{Acknowledgements}

L.F. acknowledges the fund from National Key R\&D Program of China (No. 2022YFA1004500), the fund from Research Grants Council (RGC) of the Government of Hong Kong Special Administrative Region (HKSAR) with RGC/ECS Project (No. 26200222), the fund from Guangdong Basic and Applied Basic Research Foundation (No. 2022A1515011779), the fund from the Project of Hetao Shenzhen-Hong Kong Science and Technology Innovation Cooperation Zone (No. HZQB-KCZYB-2020083), and the fund from Key Laboratory of Computational Aerodynamics, AVIC Aerodynamics Research Institute.

\bibliographystyle{elsarticle-num}
\bibliography{refs}

\end{document}